\documentstyle[12pt]{article}
\begin{document}
\setcounter{page}{1}

\centerline{\normalsize hep-ph/9405325 \hfill OITS 543}
\centerline{\normalsize \hfill May, 1994}

\vspace{15mm}

\begin{center} {\bf QUARK MASS EFFECTS IN FERMIONIC DECAYS\\
    OF THE HIGGS BOSON IN $O(\alpha_{s}^{2})$ PERTURBATIVE QCD}\\

\vspace{1cm}
               {\bf Levan R. Surguladze} \\
\vspace{2mm}
{\it Institute of Theoretical Science, University of Oregon\\
                  Eugene, OR 97403, USA}

\vspace{3mm}
\end{center}
\begin{abstract}

The results of analytical evaluation of
$O(\alpha_s^2)$ QCD contributions due to the
nonvanishing quark masses to $\Gamma_{H\rightarrow q_f\overline{q}_f}$
are presented.
The ``triangle anomaly'' type contributions are included.
As a byproduct the $O(\alpha_s^3)$ logarithmic contributions
are evaluated.
The results are presented both
in terms of running  and pole quark masses.
The partial decay modes $H \rightarrow b\overline{b}$ and
$H \rightarrow c\overline{c}$ are considered.
The calculated corrections decrease the absolute value of large and negative
$O(\alpha_s^2)$ massless limit coefficient by $\leq 1\%$ in the
intermediate mass region and by 1\%$ - $20\% in the low mass region which,
however, is experimentally ruled out.
 The results are relevant for $H \rightarrow t\overline{t}$
decay mode for the higher Higgs mass region where the mass effects are
large and important.
The high order corrections remove a very large
discrepancy between the results for
$\Gamma_{H\rightarrow q_f\overline{q}_f}$ in terms of running
and pole quark masses almost completely and
reduce the scale dependence
from about 40\% to nearly 5\%.
The remaining theoretical uncertainties are discussed.
\end{abstract}

\vspace{1cm}

{\bf 1. Introduction}
  The discovery of the Higgs boson,
which would be a crucial confirmation of the Standard Model (SM),
is one of the primary goals of modern high energy physics
\cite{rev0}. The unsuccessful search of the Higgs particle ($H$) at LEP~100
has established the present lower bound on its mass: $M_H > 63.5$ GeV
 (95\% c.l.). On the other hand, the theory is able to provide only a
broad range of the allowed Higgs mass with the upper limit about
$9M_W - (8\pi\sqrt{2}/3G_F)^{1/2}\approx 1$ TeV (see, e.g.,
\cite{rev1}). However,
if the standard spontaneous symmetry breaking mechanism is
correct, then the SM Higgs particle is expected to show up at
LEP~200 or/and LHC (for the review of Higgs search at LEP see,
e.g., \cite{rev2}) and possibly at the upgraded Fermilab Tevatron
\cite{SMW}.

   Prior to the experimental discovery, important
constraints on the Higgs properties and on its mass can be obtained
when the loop radiative corrections are taken into account. Moreover,
in the case of discovery of the Higgs boson, the precise evaluation
of the characteristics of its
production and decay processes from the first principles
of the theory would be
important in order to distinguish the
SM Higgs from its extended theoretical versions.
(For the analysis of the loop radiative effects in the
SM Higgs phenomenology see \cite{rev1}.)

  The explicit calculations show that the high order radiative
effects are large and important. For example, the calculated QCD
corrections at $M_H \sim 120$ GeV reduce the Born result for
$H \rightarrow c\overline{c}$
by about 20\% and for the $b\overline{b}$ mode by about 42\%
(see, e.g., \cite{rev1} and references therein).
There are similar size effects for $H \rightarrow gg$ and even
larger for $pp \rightarrow H+X$.

   The leading order QCD correction to the decay width
$H \rightarrow q_f\overline{q}_f$, where $q_f$ is the quark with
flavor $f$ and mass $m_f$,
has been calculated by several groups \cite{Bra}-\cite{Dre}.
In \cite{Sak} this correction has been evaluated in the zero quark mass
limit. The $O(\alpha_s^2)$ QCD correction to the
$\Gamma_{H \rightarrow q_f\overline{q}_f}$ has been computed in \cite{MPL}
(see also \cite{PRV}) in the zero quark mass limit and the results
were presented in terms of running quark mass.

   As was mentioned, although the QCD corrections depend on
$M_H$, their impact on the decay width is very large for the wide range
of the Higgs mass. Indeed, the massless limit ($m_f/M_H \rightarrow 0$)
three-loop correction \cite{MPL,PRV}
is about 10\% of the Born result and 30\% of the
leading order QCD correction (with $\alpha_s(M_H) \approx 0.15$).
On the other hand, presently there are no analytical methods to carry out
the complete three-loop calculations involving massive particles (as an
exceptions see \cite{Khn,Sop}). However, the contributions due to the
nonvanishing
quark masses might be important for modeling the behavior near the
flavor thresholds including the $t\overline{t}$ threshold.
Moreover, the possibility exists that the large $O(\alpha_s^2)$
QCD correction \cite{MPL,PRV} is an artifact of the
limit $m_f\rightarrow 0$.

   In the present work the $O(\alpha_s^2)$
corrections $\sim m_{f}^2/M_H^2$ ($f=u,d,s,c,b,t$) to the decay width
$\Gamma_{H \rightarrow q_f\overline{q}_f}$ are calculated.
The results are obtained both in terms of running and
pole quark masses. The calculated
$\sim m_f^2/M_H^2$ corrections are not entirely negligible
in the intermediate and low Higgs mass regions. For instance, at
 $M_H=(100,40,20)$GeV the $O(\alpha_s^2)$
mass corrections  increase the previous result for
$\Gamma_{H\rightarrow b\overline{b}}$ by about (1\%,5\%,23\%).
However, the low Higgs mass region is experimentally ruled out \cite{rev2}.
On the other hand, the mass corrections for the $t\overline{t}$ decay mode
in the high Higgs mass region are very large.

\vspace{4mm}
{\bf 2. Preliminary Relations.}
   The standard $SU(2)\times U(1)$ Lagrangian density of fermion-Higgs
interaction is:
\begin{equation}
L = -g_{Y}\overline{q}_fq_fH
    =-(\sqrt{2}G_F)^{1/2}m_f\overline{q}_fq_fH
     =-(\sqrt{2}G_F)^{1/2}j_fH
\label{eq:Lagr}
\end{equation}
The two-point correlation function of the scalar currents
$j_f=m_f\overline{q}_fq_f$ has the following form:
\begin{equation}
\Pi(Q^2=-s,m_f)=i\int e^{iqx}<Tj_f(x)j_f(0)>_{0}d^4x
\label{eq:PI}
\end{equation}
The decay width can be expressed as
the imaginary part of $\Pi(s+i0,m_f)$ in the standard way:
\begin{equation}
\Gamma_{H\rightarrow q_f\overline{q}_f}
         =\frac{\sqrt{2}G_F}{M_H} Im\Pi(s+i0,m_f)\biggr|_{s=M_H^2}
\label{eq:ImPi}
\end{equation}
The total hadronic decay width will be the sum over all participating
(depending on $M_H$) quark flavors:
\begin{equation}
\Gamma_{tot}^{H\rightarrow hadrons} = \sum_{f=u,d,s,...}
                            \Gamma_{H\rightarrow q_f\overline{q}_f}
\label{eq:total}
\end{equation}
For the experimentally accessible quantity - the branching fraction
of the particular decay mode one has:
\begin{equation}
BR(H\rightarrow q_f\overline{q}_f)=
    \Gamma_{H\rightarrow q_f\overline{q}_f}/
         (\Gamma_{tot}^{H\rightarrow hadrons}
          +\Gamma^{H\rightarrow \tau^{+}\tau^{-}})
\label{eq:branch}
\end{equation}

    The full $O(\alpha_s)$ analytical result
for the decay rate of $H \rightarrow q_f\overline{q}_f$
in terms of pole quark masses looks like \cite{Bra}-\cite{Dre}:
\begin{equation}
\Gamma_{H\rightarrow q_f\overline{q}_f}
      =\frac{3\sqrt{2}G_FM_H}{8\pi}m_f^2
           \biggl(1-\frac{4m_f^2}{M_H^2}\biggr)^{\frac{3}{2}}
             \biggl[1+\frac{\alpha_s(M_H)}{\pi}
                       \delta^{(1)}(\frac{m_f^2}{M_H^2})
               +O(\alpha_s^2)\biggr]
\label{eq:2loop}
\end{equation}
where:
\begin{displaymath}
\delta^{(1)}=\frac{4}{3}\biggl[\frac{a(\beta)}{\beta}
     +\frac{3+34\beta^2-13\beta^4}{16\beta^3}\log\gamma
         +\frac{21\beta^2-3}{8\beta^2}\biggr]
\end{displaymath}
\begin{displaymath}
a(\beta)=(1+\beta^2)\biggl[4Li_2(\gamma^{-1})+2Li_2(-\gamma^{-1})
          -\log\gamma\log\frac{8\beta^2}{(1+\beta)^3}\biggr]
           -\beta\log\frac{64\beta^4}{(1-\beta^2)^3}
\end{displaymath}
\begin{displaymath}
\gamma=(1+\beta)/(1-\beta),
\hspace{5mm}
\beta=(1-4m_f^2/M_H^2)^{1/2}
\end{displaymath}
and the Spence function is defined as usual:
\begin{displaymath}
Li_2(x) = -\int_{0}^{x}dx\frac{\log(1-x)}{x}
              =\sum_{n=1}^{\infty}\frac{x^n}{n^2}
\end{displaymath}
    The expansion of the r.h.s of eq.(\ref{eq:2loop}) in a power series
in terms of small $m_f^2/M_H^2$ has the following form:
\begin{eqnarray}
\lefteqn{\Gamma_{H\rightarrow q_f\overline{q}_f}
                  =\frac{3\sqrt{2}G_FM_H}{8\pi}m_f^2
        \biggl\{\biggl(1-6\frac{m_f^2}{M_H^2}+...\biggr)} \nonumber\\
 && \quad
     \hspace{-5mm} +\frac{\alpha_s(M_H)}{\pi}
                   \biggl[3-2\log\frac{M_H^2}{m_f^2}
              -\frac{m_f^2}{M_H^2}\biggl(8-24\log\frac{M_H^2}{m_f^2}\biggr)
              +...\biggr]+O(\alpha_s^2)\biggr\}
\label{eq:2loopexpan}
\end{eqnarray}
where the period covers high order terms $\sim (m_f/M_H)^{2k}$, $k=2,3...$.

\vspace{4mm}
{\bf 3. $O(\alpha_s^2)$ QCD contribution to
$\Gamma_{H\rightarrow q_f\overline{q}_f}$.}
The expansion of
the full two-point correlation function (\ref{eq:PI}) in powers
of $m_f^2/Q^2$ in the ``deep'' Euclidean region ($Q^2 \gg m_f^2$)
has the following form:
\begin{equation}
\frac{1}{m_f^2Q^2}\Pi(Q^2,m_f,m_{v})
   =\Pi_1(Q^2)+\frac{m_f^2}{Q^2}\Pi_{m_f^2}(Q^2)
      +\sum_{v=u,d,s,c,b}\frac{m_{v}^2}{Q^2}\Pi_{m_{v}^2}(Q^2)
        +...
\label{eq:Piexpan}
\end{equation}
The last term in the above expansion is due to the certain topological
types of three-loop diagrams containing virtual fermionic loop.
In fact, in these loops
the virtual top quark can also appear. This
issue will be discussed later.

   In order to evaluate the coefficient functions on the r.h.s
of eq.(\ref{eq:Piexpan}), it is
sufficient to write the diagrammatic representation for the
$\Pi(Q^2,m_f^B,m_{v}^B)$ up to the desired level of perturbation
theory and apply the appropriate projector. To $O(\alpha_s^2)$
one has:
\begin{equation}
\Pi_{m_{f}^{2n}m_{v}^{2k}}(Q^2)
         = \frac{1}{(2n)!(2k)!}
       \biggl(\frac{d}{dm_f^B}\biggr)^{2n}
        \biggl(\frac{d}{dm_{v}^B}\biggr)^{2k}
        \biggl\{\frac{\Pi(Q^2,m_f^B,m_{v}^B)}
                     {m_f^2 Q^{2(1-n-k)}}\biggr\}_{m_f^B=m_{v}^B=0}
\label{eq:Proj}
\end{equation}
where $n,k=0,1$; $n+k\leq 1$ and superscript ``B''
denotes the bare quantities.

    The one-, two- and some of the typical three-loop diagrams
contributing to the $\Pi_{i}$ are pictured on the fig.1.

\vspace{4cm}
\begin{center}
{\bf fig.1} Feynman diagrams
contributing to the $\Gamma_{H\rightarrow q_f\overline{q}_f}$
to $O(\alpha_s^2)$.\\ The total number
of three-loop graphs is 18.
\end{center}
\vspace{2mm}

    The dimensional regularization formalism
\cite{drg} and the minimal subtraction prescription \cite{MS}
in its modified form - $\overline{MS}$ \cite{MSB} are used.
The corresponding one-, two- and three-loop diagrams have been
evaluated analytically, using the special computer program HEPLoops
\cite{HEPL}, written on the analytical programming system FORM
\cite{FORM}. More details and the graph-by-graph results
will be given elsewhere.
The graph-by-graph results were summed up with
an appropriate symmetry and gauge group weights. In the obtained
expression for the $\Pi_{i}$ in terms of bare quantities
one renormalizes the coupling and the quark mass factors:
\begin{equation}
\alpha_s^{B}=\alpha_s\biggl[1-\frac{\alpha_s}{4\pi}
        \biggl(\frac{11}{3}C_A-\frac{4}{3}TN\biggr)\frac{1}{\varepsilon}
        \biggr]
\label{eq:Asren}
\end{equation}
\begin{equation}
(m_f^B)^2 =
       m_f^2\biggl\{1-\frac{\alpha_s}{4\pi}\frac{6C_F}{\varepsilon}
             +\biggl(\frac{\alpha_s}{4\pi}\biggr)^2C_F
      \biggl[(11C_A+18C_F-4TN)\frac{1}{\varepsilon^2}
    -\biggl(\frac{97}{6}C_A+\frac{3}{2}C_F-\frac{10}{3}TN\biggr)
              \frac{1}{\varepsilon}\biggr]\biggr\}
\label{eq:mren}
\end{equation}
where, $\varepsilon=(4-D)/2$ is the parameter of
dimensional regularization, $D$ is the dimension of space-time;
$N$ is the number of participating quark flavors.
The eigenvalues of Casimir operators for the adjoint and the
fundamental representations of $SU_c(3)$ gauge group
are  $C_A=3$, $C_F=4/3$ correspondingly and $T=1/2$.

    The obtained expressions for $\Pi_i$ at each order of $\alpha_s$
are the polynomials with respect to $1/\varepsilon$ and
$\log\mu_{\overline{MS}}^2/Q^2$. However, there are no terms like
$(1/\varepsilon^n)(\log\mu_{\overline{MS}}^2/Q^2)^k$.
They will appear only
at higher orders $\sim m_f^2 m_f^4/Q^4$ and represent the
infrared mass logarithms. The poles can be removed by an
additive renormalization. The
imaginary part of each of the $\Pi_i$ is finite after the
regularization is removed. Here the usual way \cite{MPL,PRV}
of the introduction of Adler $D$-function \cite{Adl} (the
$\Pi$-function differentiated with respect of $Q^2$) which
is finite and renormalization group (RG) invariant, can be avoided.
Instead, one can operate directly with the correlation function $\Pi_i$.
Namely, one should analytically continue it from Euclidean to
Minkowski space and take the imaginary part at
$s=M_H^2$ ( eq.(\ref{eq:ImPi}) ). One obtains the
following analytical result for the standard QCD with $SU_c(3)$
gauge group:
\begin{eqnarray}
\lefteqn{\Gamma_{H\rightarrow q_f\overline{q}_f}
                  =\frac{3\sqrt{2}G_FM_H}{8\pi}m_f^2\biggl\{
  1+\frac{\alpha_s}{\pi}\biggl(\frac{17}{3}
+2\log\frac{\mu_{\overline{MS}}^2}{M_H^2}\biggr)}\nonumber\\
 && \quad
   +\biggl(\frac{\alpha_s}{\pi}\biggr)^2
      \biggl[\frac{10801}{144}-\frac{19}{2}\zeta(2)-\frac{39}{2}\zeta(3)
         +\frac{106}{3}\log\frac{\mu_{\overline{MS}}^2}{M_H^2}
         +\frac{19}{4}\log^2\frac{\mu_{\overline{MS}}^2}{M_H^2}\nonumber\\
 && \quad \hspace{1cm}
  -N\biggl(\frac{65}{24}-\frac{1}{3}\zeta(2)-\frac{2}{3}\zeta(3)
         +\frac{11}{9}\log\frac{\mu_{\overline{MS}}^2}{M_H^2}
         +\frac{1}{6}\log^2\frac{\mu_{\overline{MS}}^2}{M_H^2}\biggr)\biggr]
                                                               \nonumber\\
 && \quad \hspace{35mm}
     -\frac{m_f^2}{M_H^2}\biggl<
        6+\frac{\alpha_s}{\pi}\biggl(40
                +24\log\frac{\mu_{\overline{MS}}^2}{M_H^2}\biggr)\nonumber\\
 && \quad
   +\biggl(\frac{\alpha_s}{\pi}\biggr)^2
      \biggl[\frac{2383}{4}-162\zeta(2)-166\zeta(3)
         +371\log\frac{\mu_{\overline{MS}}^2}{M_H^2}
         +81\log^2\frac{\mu_{\overline{MS}}^2}{M_H^2}\nonumber\\
 && \quad \hspace{1cm}
  -N\biggl(\frac{313}{18}-4\zeta(2)-4\zeta(3)
         +10\log\frac{\mu_{\overline{MS}}^2}{M_H^2}
         +2\log^2\frac{\mu_{\overline{MS}}^2}{M_H^2}\biggr)\biggr]\biggr>
                                                               \nonumber\\
 && \quad
    +\biggl(\frac{\alpha_s}{\pi}\biggr)^2
     \sum_{v=u,d,s,c,b}\frac{m_{v}^2}{M_H^2}12 \biggr\}
\label{eq:gamma0}
\end{eqnarray}
where the Riemann function $\zeta(2)=\pi^2/6$ arose
from the analytical continuation of $\log^3\mu_{\overline{MS}}^2/Q^2$
terms and $\zeta(3)=1.202056903$.
The last term in eq.(\ref{eq:gamma0}) represents the contributions
from the three-loop diagrams in fig.1 containing the virtual quark loop
and the
``triangle anomaly'' type corrections from the graphs in fig.2. Their
contribution vanishes in the massless quark limit.

\vspace{4cm}
\begin{center}
{\bf fig.2} The ``triangle anomaly'' type diagrams.
\end{center}
\vspace{2mm}

      The leading three-loop term
$\sim m_f^2$ (the massless approximation)
coincides with the one obtained in \cite{MPL,PRV}, while the three-loop
results $\sim m_f^4, m_f^2m_{v}^2$ are new.

\vspace{4mm}
{\bf 4. Renormalization group analysis.}
An observable quantity, in particular the calculated decay width
is invariant under the RG transformations and
obeys the homogeneous RG-equation:
\begin{equation}
\biggl(\mu^2\frac{\partial}{\partial\mu^2}
   +\beta(\alpha_s)\alpha_s\frac{\partial}{\partial\alpha_s}
   -\gamma_m(\alpha_s)\sum_{l=f,v}
        m_l\frac{\partial}{\partial m_l} \biggr)
    \Gamma_{H\rightarrow q_f\overline{q}_f}
            \biggl(\frac{\mu^2}{M_H^2},m_f,m_{v},\alpha_s\biggr)=0
\label{eq:RG}
\end{equation}
where the QCD $\beta$-function and the mass anomalous dimension are
introduced in the standard way:
\begin{equation}
\mu^2\frac{d\alpha_s}{d\mu^2}=\alpha_s\beta(\alpha_s)
     =-\alpha_s\sum_{k\geq 0}\beta_k\biggl(\frac{\alpha_s}{\pi}\biggr)^{k+1}
\label{eq:beta}
\end{equation}
\begin{equation}
-\frac{\mu^2}{m_f}\frac{d m_f}{d\mu^2}=\gamma_m(\alpha_s)
     =\sum_{k\geq 0}\gamma_k\biggl(\frac{\alpha_s}{\pi}\biggr)^{k+1}
\label{eq:gammam}
\end{equation}
The known $\beta$-function coefficients are \cite{3lb}:
\begin{equation}
\beta_{0}=\frac{1}{4}\biggl(11-\frac{2}{3}N\biggr), \hspace{3mm}
\beta_{1}=\frac{1}{16}\biggl(102-\frac{38}{3}N\biggr), \hspace{3mm}
\beta_{2}=\frac{1}{64}\biggl(\frac{2857}{2}-\frac{5033}{18}N
                +\frac{325}{54}N^2\biggr)
\label{eq:3lbeta}
\end{equation}
and for the $\gamma_m(\alpha_s)$ one has \cite{3lg}:
\begin{equation}
\gamma_{0}=1, \hspace{3mm}
\gamma_{1}=\frac{1}{16}\biggl(\frac{202}{3}-\frac{20}{9}N\biggr), \hspace{3mm}
\gamma_{2}=\frac{1}{64}\biggl[1249-\biggl(\frac{2216}{27}+\frac{160}{3}
             \zeta(3)\biggr)N
                -\frac{140}{81}N^2\biggr].
\label{eq:3lgamma}
\end{equation}

   The solution of the RG eq.(\ref{eq:RG}) can conveniently be written
as follows:
\begin{equation}
\Gamma_{H\rightarrow q_f\overline{q}_f}
      = \Gamma_{0}m_f^2(\mu)\sum_{0\leq j \leq i}
    \biggl(\frac{\alpha_s(\mu)}{\pi}\biggr)^i
        \log^j\frac{\mu^2}{M_H^2}
      \biggl(a_{ij}+\frac{m_f^2(\mu)}{M_H^2}b_{ij}
      +\sum_{v}\frac{m_{v}^2(\mu)}{M_H^2}c_{ij}\biggr)
\label{eq:gamma0gen}
\end{equation}
where the coefficients $a_{ij}$, $b_{ij}$ and $c_{ij}$ are the same
as the ones in eq.(\ref{eq:gamma0}).
Applying the differential operator $\mu^2 d/d\mu^2$ to
the both sides of the eq.(\ref{eq:gamma0gen}) and taking into account
the RG-invariance of $\Gamma_{H\rightarrow q_f\overline{q}_f}$ and the
eqs. (\ref{eq:beta},\ref{eq:gammam}), one obtains at the $O(\alpha_s)$:
\begin{equation}
a_{11}=2\gamma_0a_{00}, \hspace{3mm} b_{11}=4\gamma_0b_{00},
\label{eq:relations1}
\end{equation}
at the $O(\alpha_s^2)$:
\begin{displaymath}
a_{21}=2\gamma_1a_{00}+(\beta_0+2\gamma_0)a_{10},
\end{displaymath}
\begin{displaymath}
a_{22}=(\beta_0+2\gamma_0)a_{11}/2=(\beta_0+2\gamma_0)\gamma_0a_{00},
\end{displaymath}
\begin{displaymath}
b_{21}=4\gamma_1b_{00}+(\beta_0+4\gamma_0)b_{10},
\end{displaymath}
\begin{equation}
b_{22}=(\beta_0+4\gamma_0)b_{11}/2=2(\beta_0+4\gamma_0)\gamma_0b_{00},
\label{eq:relations2}
\end{equation}
and at the $O(\alpha_s^3)$:
\begin{displaymath}
a_{31}=2(\beta_0+\gamma_0)a_{20}+(\beta_1+2\gamma_1)a_{10}+2\gamma_2a_{00},
\end{displaymath}
\begin{displaymath}
a_{32}=(\beta_0+\gamma_0)a_{21}+(\beta_1+2\gamma_1)a_{11}/2
      =(\beta_0+\gamma_0)[2\gamma_1a_{00}+(\beta_0+2\gamma_0)a_{10}]
       +(\beta_1+2\gamma_1)\gamma_0a_{00},
\end{displaymath}
\begin{displaymath}
a_{33}=2(\beta_0+\gamma_0)a_{22}/3=2\gamma_0
         (\beta_0+\gamma_0)(\beta_0+2\gamma_0)a_{00}/3,
\end{displaymath}
\begin{displaymath}
b_{31}=2(\beta_0+2\gamma_0)b_{20}+(\beta_1+4\gamma_1)b_{10}+4\gamma_2b_{00},
\end{displaymath}
\begin{displaymath}
b_{32}=(\beta_0+2\gamma_0)b_{21}+(\beta_1+4\gamma_1)\frac{b_{11}}{2}
      =(\beta_0+2\gamma_0)[4\gamma_1b_{00}+(\beta_0+4\gamma_0)b_{10}]
       +(\beta_1+4\gamma_1)2\gamma_0b_{00},
\end{displaymath}
\begin{displaymath}
b_{33}=2(\beta_0+2\gamma_0)b_{22}/3=4\gamma_0
         (\beta_0+2\gamma_0)(\beta_0+4\gamma_0)b_{00}/3.
\end{displaymath}
\begin{equation}
c_{31}=2(\beta_0+2\gamma_0)c_{20}
\label{eq:relations3}
\end{equation}
Since the diagrams with the virtual fermionic loop first appear at
$O(\alpha_s^2)$, $c_{00}=c_{1j}=c_{21}=c_{22}=0$.
   The relations (\ref{eq:relations1},\ref{eq:relations2}) provide a
powerful check of the $O(\alpha_s^2)$ calculations, while the relations
(\ref{eq:relations3}) allow one to evaluate the
$\log$-terms at $O(\alpha_s^3)$,
without explicit calculations of the corresponding four-loop diagrams.
With those relations, the information available at present, namely the
QCD $\beta$-function, mass anomalous dimension and the
2-point correlation function up to the three-loop level, is fully exploited.
In fact, the similar relations can be derived for the correlation function
$\Pi$. However, the RG-equation for $\Pi$ is not a homogeneous one and the
anomalous dimension function up to the corresponding order of $\alpha_s$
is needed.

   The solution of the RG-equation (\ref{eq:RG})
at $\mu^2_{\overline{MS}}=M_H^2$ has the following form:
\begin{eqnarray}
\lefteqn{\Gamma_{H\rightarrow q_f\overline{q}_f}
     =\frac{3\sqrt{2}G_FM_H}{8\pi}m_f^2(M_H)\biggl\{
                1-6\frac{m_f^2(M_H)}{M_H^2}
  +\frac{\alpha_s(M_H)}{\pi}\biggl(5.66667-40\frac{m_f^2(M_H)}{M_H^2}\biggr)}
                                                             \nonumber\\
 && \quad
  +\biggl(\frac{\alpha_s(M_H)}{\pi}\biggr)^2
     \biggl[35.93996-1.35865N
        -\frac{m_f^2(M_H)}{M_H^2}\biggl(129.72924-6.00093N\biggr)
                                                              \nonumber\\
 && \quad \hspace{3cm}
            +12\sum_{v=u,d,s,c,b}m_{v}^2(M_H)/M_H^2\biggr]\biggr\}.
\label{eq:RGimpr}
\end{eqnarray}
The running coupling is parametrized as follows:
\begin{equation}
\frac{\alpha_s(M_H^2)}{\pi}=\frac{1}{\beta_0 L}-\frac{\beta_1 \log L}
{\beta_0^3 L^2}+\frac{1}{\beta_0^5 L^3}(\beta_1^2 \log^2 L-\beta_1^2 \log L
+\beta_2 \beta_0-\beta_{1}^{2})+O(L^{-4}),
\label{eq:Asparametr}
\end{equation}
where $L=\log(M_H^2/\Lambda_{\overline{MS}}^2)$.
For the running mass one has:
\begin{equation}
m_f(\mu_1)/m_f(\mu_2)=\phi(\alpha_s(\mu_1))/\phi(\alpha_s(\mu_2)),
\label{eq:mrun}
\end{equation}
where,
\begin{eqnarray}
\lefteqn{\phi(\alpha_s(\mu))=\biggl(2\beta_0\frac{\alpha_s(\mu)}{\pi}\biggr)
                              ^{\frac{\gamma_0}{\beta_0}}
       \biggl\{1
     +\biggl(\frac{\gamma_1}{\beta_0}
         -\frac{\beta_1\gamma_0}{\beta_0^2}\biggr)\frac{\alpha_s(\mu)}{\pi}}
                                                             \nonumber\\
 && \quad \hspace{-3mm}
     +\frac{1}{2}\biggl[\biggl(\frac{\gamma_1}{\beta_0}
         -\frac{\beta_1\gamma_0}{\beta_0^2}\biggr)^2
       +\frac{\gamma_2}{\beta_0}
       -\frac{\beta_1\gamma_1}{\beta_0^2}-\frac{\beta_2\gamma_0}{\beta_0^2}
       +\frac{\beta_1^2\gamma_0}{\beta_0^3}
         \biggr]\biggl(\frac{\alpha_s(\mu)}{\pi}\biggr)^2\biggr\}
\label{eq:f}
\end{eqnarray}
In the eqs. (\ref{eq:RGimpr})-(\ref{eq:f}) all appropriate quantities
are evaluated for the $N$ active quark flavors. $N$ can be determined
according to the scale of $M_H$. At present one usually considers $N=5$.

\vspace{4mm}
{\bf 5. $\Gamma_{H\rightarrow q_f\overline{q}_f}$ in terms of pole quark mass.}
For the heavy flavor decay mode of the Higgs, it is relevant
to parametrize the decay rate in terms of
pole quark mass (see, e.g., \cite{rev1}). Below the result (\ref{eq:RGimpr})
will be rewritten in terms of pole quark mass,
assuming that heavy quark is not exactly on-shell.

   First, one derives the following general evolution equation
for the running coupling to $O(\alpha_s^3)$:
\begin{eqnarray}
\lefteqn{\hspace{-25mm}\frac{\alpha_s^{(n)}(\mu)}{\pi}
                         =\frac{\alpha_s^{(N)}(M)}{\pi}
      +\biggl(\frac{\alpha_s^{(N)}(M)}{\pi}\biggr)^2
      \biggl(\beta_0^{(N)}\log\frac{M^2}{\mu^2}
      +\frac{1}{6}\sum_{l}\log\frac{m_l^2}{\mu^2} \biggr)
      +\biggl(\frac{\alpha_s^{(N)}(M)}{\pi}\biggr)^3
      \biggl[\beta_1^{(N)}\log\frac{M^2}{\mu^2}}
                                                             \nonumber\\
 &&
      +\frac{19}{24}\sum_{l}\log\frac{m_l^2}{\mu^2}
      +\biggl(\beta_0^{(N)}\log\frac{M^2}{\mu^2}
      +\frac{1}{6}\sum_{l}\log\frac{m_l^2}{\mu^2} \biggr)^2
                           -\frac{25}{72}(N-n)\biggr]
\label{eq:Astransform}
\end{eqnarray}

where the superscript $n$ ($N$) indicates that the corresponding quantity
is evaluated for $n$ ($N$) numbers of
participating quark flavors.
Conventionally (see, e.g., \cite{MAR}), $n$ ($N$) is specified to be
the number
of quark flavors with mass $\leq \mu$ ($\leq M$). However, the
eq.(\ref{eq:Astransform}) is relevant for any $n\leq N$ and arbitrary
$\mu$ and $M$, regardless the conventional specification of the number of
quark flavors. The $\log m_l/\mu$
terms are due to the ``quark threshold'' crossing effects and the constant
coefficients $1/6=\beta_0^{(k-1)}-\beta_0^{(k)}$,
$19/24=\beta_1^{(k-1)}-\beta_1^{(k)}$ represent
the contributions of the quark loop in the $\beta$-function.
The sum runs over $N-n$ quark flavors (e.g., $l=b$ if $n=4$ and $N=5$).
Note that $m_l$ is the pole mass of the quark with flavor $l$.
For the on-shell definition of the quark masses the eq.(\ref{eq:Astransform})
changes - the constant $-25/72$ should be substituted by $+7/72$.
The above equation is derived based on the eq.(\ref{eq:Asparametr}),
the QCD matching conditions for $\alpha_s$ at ``quark thresholds''
\cite{BHS,BeW} and the one-loop relation between on-shell and
pole quark masses.
The eq.(\ref{eq:Astransform}) is consistent with QCD matching relation
at $m_f(m_f)$ \cite{BeW} (see also \cite{SAN}):
\begin{equation}
\alpha_s^{(N_f-1)}(m_f(m_f))=\alpha_s^{(N_f)}(m_f(m_f))
               +(\alpha_s^{(N_f)}(m_f(m_f)))^3(C_A/9-17C_F/96)/\pi^2
\label{eq:2lmatch}
\end{equation}
Here and below $N_f$ is the number of quark flavors $u,d,...,f$.
Note that the nonlogarithmic constant at $O(\alpha_s^3)$ in
eq.(\ref{eq:Astransform}) will not contribute in further analysis.

    Next, using the scaling properties of the $MS$ running mass
and the eq.(\ref{eq:Astransform}), one obtains
the following matching condition:
\begin{eqnarray}
\lefteqn{\hspace{-12mm}m_f^{(N-1)}(\mu)=m_f^{(N)}(\mu)\biggl\{1+
   \biggl(\frac{\alpha_s^{(N)}(\mu)}{\pi}\biggr)^2
   \biggl[\delta(m_f,m_{f'})-\frac{5}{36}\log\frac{\mu^2}{m_f^2}
   -\frac{1}{12}\log^2\frac{\mu^2}{m_f^2}}
                                                               \nonumber\\
 && \quad \hspace{65mm}
    +\frac{1}{6}\log\frac{\mu^2}{m_f^2}\log\frac{\mu^2}{m_{f'}^2}
    -\frac{2}{9}\log\frac{m_{f'}^2}{m_f^2}\biggr]\biggr\}
\label{eq:massmatch}
\end{eqnarray}
where the constant terms are:
$1/12=\gamma_0(\beta_0^{(k-1)}-\beta_0^{(k)})/2$,
$5/36=\gamma_1^{(k-1)}-\gamma_1^{(k)}$ and
$2/9=C_F(\beta_0^{(k-1)}-\beta_0^{(k)})$.
In general the $\delta(m_f,m_{f'})$ is the
finite contribution of the single virtual heavier
quark with mass $m_{f'}$, entering when one increases the
number of flavors
from $N-1$ to $N$ (one can also consider the particular case
$m_{f'}=m_f$).
{}From the two-loop on-shell quark mass
renormalization one has \cite{BRH}:
\begin{equation}
\delta(m_f,m_{f'})=-\zeta(2)/3-71/144
    +(4/3)\Delta(m_{f'}/m_f)
\label{eq:deltaM}
\end{equation}
where
\begin{equation}
\Delta(r)=\frac{1}{4}\biggl[\log^2r+\zeta(2)-\biggl(\log r
        +\frac{3}{2}\biggr)r^2
        -(1+r)(1+r^3)L_{+}(r)-(1-r)(1-r^3)L_{-}(r)\biggr],
\label{eq:delta}
\end{equation}
\begin{displaymath}
L_{\pm}(r) = \int_{0}^{1/r}dx\frac{\log x}{x \pm 1}.
\end{displaymath}
$L_{\pm}(r)$ can be evaluated for different quark mass ratios
$r$ numerically (table 1).

      One can relate the $\overline{MS}$ quark mass $m_f(m_f)$
to the pole mass $m_f$ using the $O(\alpha_s^2)$ on-shell results of
\cite{BRH}:
\begin{equation}
m_f^{(N_f)}(m_f)=m_f[1-4\alpha_s^{(N_f)}(m_f)/3\pi
     +(16/9-K_{f})(\alpha_s^{(N_f)}(m_f)/\pi)^2],
\label{eq:mtopole}
\end{equation}
where
\begin{equation}
K_f= \frac{3817}{288}+\frac{2}{3}(2+\log2)\zeta(2)-\frac{1}{6}\zeta(3)
  -\frac{N_f}{3}\biggl(\zeta(2)+\frac{71}{48}\biggr)
  +\frac{4}{3}\sum_{m_l \leq m_f} \Delta\biggl(\frac{m_l}{m_f}\biggr).
\label{eq:K}
\end{equation}
The first four terms in $K_f$ represent the QCD contribution with $N_f$
massless quarks, while the sum is the correction due to the
$N_f$ nonvanishing quark masses.

  Combining the eqs. (\ref{eq:mrun},\ref{eq:f}) and the eqs.
(\ref{eq:Astransform})-(\ref{eq:mtopole}), one obtains the
relation between the $\overline{MS}$ quark mass $m_f(M_H)$ renormalized
at $M_H$ and evaluated for the $N$-flavor theory and the pole quark
mass $m_f$:
\begin{eqnarray}
\lefteqn{\hspace{-2cm}m_f^{(N)}(M_H)=m_f\biggl\{1
   -\frac{\alpha_s^{(N)}(M_H)}{\pi}
      \biggl(\frac{4}{3}+\gamma_0\log\frac{M_H^2}{m_f^2}\biggr)
      -\biggl(\frac{\alpha_s^{(N)}(M_H)}{\pi}\biggr)^2\biggl[K_f
        +\sum_{m_f<m_{f'}<M_H}\delta(m_f,m_{f'})}
                                                                \nonumber\\
 && \quad
        -\frac{16}{9}
        +\biggl(\gamma_1^{(N)}-\frac{4}{3}\gamma_0
          +\frac{4}{3}\beta_0^{(N)}\biggr)\log\frac{M_H^2}{m_f^2}
      +\frac{\gamma_0}{2}(\beta_0^{(N)}-\gamma_0)\log^2\frac{M_H^2}{m_f^2}
                                                     \biggr]\biggr\}
\label{eq:mMHtopole}
\end{eqnarray}
Note that $N$ is specified according to the size of $M_H$ and has
no correlation with the quark mass $m_f$. Thus, for instance, one can
apply the eq.(\ref{eq:mMHtopole}) to the charm mass $m_c^{(5)}(M_H)$
evaluated for five-flavor
theory.

    Substituting eqs.
(\ref{eq:mMHtopole},\ref{eq:K},\ref{eq:3lgamma},\ref{eq:3lbeta})
into the eq.(\ref{eq:RGimpr}),
one obtains  the general form for the decay rate
 $\Gamma_{H\rightarrow q_f\overline{q}_f}$ in terms of the pole quark masses:
\begin{eqnarray}
\lefteqn{\Gamma_{H\rightarrow q_f\overline{q}_f}
     =\frac{3\sqrt{2}G_FM_H}{8\pi}m_f^2\biggl\{1-6\frac{m_f^2}{M_H^2}
  +\frac{\alpha_s^{(N)}(M_H)}{\pi}
   \biggl[3-2\log\frac{M_H^2}{m_f^2}
     -\frac{m_f^2}{M_H^2}\biggl(8-24\log\frac{M_H^2}{m_f^2}\biggr)\biggr]}
                                                             \nonumber\\
 && \quad \hspace{-1cm}
  +\biggl(\frac{\alpha_s^{(N)}(M_H)}{\pi}\biggr)^2
   \biggl<\frac{697}{18}-\biggl(\frac{73}{6}+\frac{4}{3}\log 2\biggr)\zeta(2)
          -\frac{115}{6}\zeta(3)
          -N\biggl(\frac{31}{18}-\zeta(2)-\frac{2}{3}\zeta(3)\biggr)
                                                             \nonumber\\
 && \quad
    -\frac{m_f^2}{M_H^2}\biggl[171-(194+16\log2)\zeta(2)-162\zeta(3)
          -N\biggl(\frac{50}{9}-12\zeta(2)-4\zeta(3)\biggr)\biggr]
                                                             \nonumber\\
 && \quad
    -\biggl[\frac{87}{4}-\frac{13}{18}N
            -\frac{m_f^2}{M_H^2}\biggl(221-\frac{26}{3}N\biggr)\biggr]
                \log\frac{M_H^2}{m_f^2}
    -\biggl[\frac{3}{4}-\frac{1}{6}N
            +\frac{m_f^2}{M_H^2}(15+2N)\biggr]\log^2\frac{M_H^2}{m_f^2}
                                                             \nonumber\\
 && \quad
    -\biggl(\frac{8}{3}-32\frac{m_f^2}{M_H^2}\biggr)
        \sum_{m_l<M_H}\Delta\biggl(\frac{m_l}{m_f}\biggr)
      +12\sum_{m_{v}<M_H}\frac{m_{v}^2}{M_H^2} \biggr>\biggr\}
\label{eq:Polemassresf}
\end{eqnarray}
 The above result
confirms the asymptotic form (\ref{eq:2loopexpan}) of the two-loop
exact result (\ref{eq:2loop}), while the $O(\alpha_s^2)$ expression is new.

The numerical values of $\Delta(m_l/m_f)$
defined in the eq.(\ref{eq:delta}) are given in the table 1.
Quark masses can be estimated from QCD sum rules:
$m_b=4.72$GeV \cite{mb}, $m_c=1.46$GeV \cite{mc}, $m_s=0.27$GeV
\cite{ms}, $m_u+m_d \approx m_s/13 \approx 0.02$GeV \cite{mud}.
For the upper bound of the top mass the latest CDF result
$m_t \leq 174+10+13$ GeV \cite{CDF} is used.
On the other hand, for the lower bound more conservative value $m_t>131$ GeV
given by D0 \cite{D0} is used. The values of $\Delta(m_l/m_t)$ are given
in the table 1 for both the lower and the upper bounds of the top mass.
\begin{center}
Table 1. The numerical values of $\Delta(m_l/m_f)$.\\

\vspace{2mm}

\begin{tabular}{|l|l|r|}    \hline
$f$-flavor  & $l$-flavor & $\Delta(m_l/m_f)$ \\ \hline
$t$         &  $u$\hspace{1mm}or\hspace{1mm}$d$  & $\leq$ 0.0001       \\
$t$         &  $s$         & 0.0025-0.0017     \\
$t$         &  $c$         & 0.0137-0.0091     \\
$t$         &  $b$         & 0.0435-0.0291     \\
$t$         &  $t$         &        0.8587     \\
$b$         &  $u$\hspace{1mm}or\hspace{1mm}$d$  & 0.0027            \\
$b$         &  $s$         & 0.0673            \\
$b$         &  $c$         & 0.3290            \\
$b$         &  $b$         & 0.8587            \\
$c$         &  $u$\hspace{1mm}or\hspace{1mm}$d$  & 0.0084            \\
$c$         &  $s$         & 0.2045            \\
$c$         &  $c$         & 0.8587            \\
$c$         &  $b$         & 1.9015            \\
$s$         &  $u$\hspace{1mm}or\hspace{1mm}$d$  & 0.0454            \\
$s$         &  $s$         & 0.8587            \\
$s$         &  $c$         & 2.5767            \\
$s$         &  $b$         & 4.5605            \\
\hline
\end{tabular}
\end{center}
\vspace{2mm}

    In order to achieve better numerical precision within the wider
range of $M_H$ it is more appropriate to use
full one- and two-loop results (eq.(\ref{eq:2loop}))
 and the mass corrected three-loop result given by the
eq.(\ref{eq:Polemassresf}).
For the dominant decay mode
$H\rightarrow b\overline{b}$ with five participating quark flavors one
obtains:
\begin{eqnarray}
\lefteqn{\Gamma_{H\rightarrow b\overline{b}}
     =\frac{3\sqrt{2}G_FM_H}{8\pi}m_b^2
     \biggl\{\biggl(1-\frac{4m_b^2}{M_H^2}\biggr)^{\frac{3}{2}}
  +\frac{\alpha_s^{(5)}(M_H)}{\pi}\delta^{(1)}\biggl(\frac{m_b^2}{M_H^2}\biggr)
       \biggl(1-\frac{4m_b^2}{M_H^2}\biggr)^{\frac{3}{2}} }
                                                             \nonumber\\
 && \quad
  +\biggl(\frac{\alpha_s^{(5)}(M_H)}{\pi}\biggr)^2
   \biggl[-2.23039+266.13393\frac{m_b^2}{M_H^2}
                                                             \nonumber\\
 && \quad
     -\biggl(18.13889-177.66667\frac{m_b^2}{M_H^2}\biggr)
                        \log\frac{M_H^2}{m_b^2}
     +\biggl(0.08333-25\frac{m_b^2}{M_H^2}\biggr)
                        \log^2\frac{M_H^2}{m_b^2}
                                                              \nonumber\\
  && \quad  \hspace{15mm}
          -\biggl(2.66667-32\frac{m_b^2}{M_H^2}\biggr)
         \sum_{m_l\leq m_b}\Delta\biggl(\frac{m_l}{m_b}\biggr)
           +12\sum_{m_{v}\leq m_b}\frac{m_{v}^2}{M_H^2} \biggr]\biggr\}
\label{eq:Polemassresb}
\end{eqnarray}
where $\delta^{(1)}$ is defined in eq.(\ref{eq:2loop}).

   At the present
LEP lower bound on Higgs mass $M_H\approx60$GeV one has:
\begin{equation}
\Gamma_{H\rightarrow b\overline{b}}
    =\frac{3\sqrt{2}G_FM_H}{8\pi}m_b^2\biggl\{(1-0.037)
     +\frac{\alpha_s^{(5)}(M_H)}{\pi}(-7.17+0.71)
     +\biggl(\frac{\alpha_s^{(5)}(M_H)}{\pi}\biggr)^2
       (-95.67+3.57)\biggr\}
\label{eq:numer}
\end{equation}
where the second number in each bracket corresponds to the
leading order quark mass
correction, while the first one is the massless contribution.
One can see that the mass corrections
always decrease the large massless contribution.
For the lower values of $M_H$ the relative mass corrections increase
rapidly. For example, at $M_H=40$GeV the $O(\alpha_s^2)$ mass correction
decrease the large and negative massless coefficient by more then 10\%
and correspondingly increase the decay rate by about 5\%.

  Finally, one also needs
to include weak and electromagnetic contributions \cite{Kni,rev1}.
However, they will not affect further analysis here.

The dependence of the partial decay rate
$\Gamma_{H\rightarrow b\overline{b}}$ on the mass of the Higgs particle
$M_H$ is plotted in fig.3.

\vspace{8cm}
\begin{center}
{\bf fig.3} The partial decay rate
              $\Gamma_{H\rightarrow b\overline{b}}$ vs. Higgs mass
\end{center}
\vspace{1mm}

In the region of $M_H$ shown in the fig.3 the three-loop
mass corrections are small and does not affect the overal picture.
However, in the low mass region their impact is significant (fig.4).

   To obtain the decay rate $\Gamma_{H\rightarrow c\overline{c}}$
one should substitute $b\rightarrow c$ everywhere in
eq.(\ref{eq:Polemassresb}) except the summation bounds, which remain the same.
The $b$-quark ``threshold'' crossing effect will be represented
by the term $\sim \Delta(m_b/m_c)$.
\newpage
{}.
\vspace{65mm}
\begin{center}
{\bf fig.4} The partial decay rate
              $\Gamma_{H\rightarrow b\overline{b}}$ vs. Higgs mass
               (low mass region)
\end{center}

    Since the derivation of the eqs. (\ref{eq:RGimpr},\ref{eq:Polemassresf})
has not been restricted by the particular quark flavor, one could apply
these results to the $H \rightarrow t\overline{t}$ decay channel.
\footnote{The author is grateful to the referee for Physics Letters B
for his suggestion.} The only restriction that apply is that
the Higgs mass $M_H$ must be sufficiently larger then $2m_t$ in order the
process to be allowed and the expansion in terms of $m_t^2/M_H^2$ to be
a legitimate.
Note, that in those equations for the $t\overline{t}$ mode one should consider
six flavor theory and the summation index $v$ will run over all six quark
flavors. Thus, from the eq. (\ref{eq:Polemassresf}) at $m_f=m_t$ and
$N_f=6$ one obtains:
\begin{eqnarray}
\lefteqn{\Gamma_{H\rightarrow t\overline{t}}
     =\frac{3\sqrt{2}G_FM_H}{8\pi}m_t^2
     \biggl\{\biggl(1-\frac{4m_t^2}{M_H^2}\biggr)^{\frac{3}{2}}
  +\frac{\alpha_s^{(6)}(M_H)}{\pi}\delta^{(1)}\biggl(\frac{m_t^2}{M_H^2}\biggr)
       \biggl(1-\frac{4m_t^2}{M_H^2}\biggr)^{\frac{3}{2}} }
                                                             \nonumber\\
 && \quad
  +\biggl(\frac{\alpha_s^{(6)}(M_H)}{\pi}\biggr)^2
   \biggl[-1.50631+247.14204\frac{m_t^2}{M_H^2}
                                                             \nonumber\\
 && \quad
     -\biggl(17.41667-169\frac{m_t^2}{M_H^2}\biggr)
                        \log\frac{M_H^2}{m_t^2}
     +\biggl(0.25-27\frac{m_t^2}{M_H^2}\biggr)
                        \log^2\frac{M_H^2}{m_t^2}
                                                              \nonumber\\
  && \quad  \hspace{15mm}
          -\biggl(2.66667-32\frac{m_t^2}{M_H^2}\biggr)
         \sum_{m_l\leq m_t}\Delta\biggl(\frac{m_l}{m_t}\biggr)
           +12\sum_{m_{v}\leq m_t}\frac{m_{v}^2}{M_H^2} \biggr]\biggr\}
\label{eq:Polemassrest}
\end{eqnarray}
The numerical values of $\Delta(m_l/m_t)$
defined in the eq.(\ref{eq:delta}) are given in the table 1.
In the above equation the known exact dependence on the top mass at
one and two-loop levels is used (see eq.(\ref{eq:2loop})).
It is straightforward to express the $\alpha_s^{(6)}$ in terms
of $\alpha_s^{(5)}$ using the relation (\ref{eq:Astransform}).
For the conservative lower bound
on the top mass $m_t=131$ GeV and for the Higgs mass at, for instance,
$M_H=400$ GeV one obtains:
\newpage
\begin{eqnarray}
\lefteqn{\Gamma_{H\rightarrow t\overline{t}}
    =\frac{3\sqrt{2}G_FM_H}{8\pi}m_t^2\biggl\{
        (1-0.644+0.069+0.005+0.0008+...) } \nonumber\\
 && \hspace{11mm}
     +\frac{\alpha_s^{(6)}(M_H)}{\pi}
        (-1.465+4.889-1.739-0.005-0.0003+...) \nonumber\\
 &&  \hspace{10mm}
     +\biggl(\frac{\alpha_s^{(6)}(M_H)}{\pi}\biggr)^2
       (-41.59+57.02+...)\biggr\}
\label{eq:numert}
\end{eqnarray}
In the above equation at each order of $\alpha_s^{(6)}$ the expansion in
$m_t^2/M_H^2$ (at $m_t=131$ GeV and $M_H=400$ GeV) is given.
One can see that for the case
$H \rightarrow t\overline{t}$ the mass corrections are important. In fact,
the massless limit coefficients (e.g., at $O(\alpha_s)$) does not reproduce
even a correct sign.

\vspace{3mm}

{\bf 6. Scheme-scale dependence and the theoretical uncertainties.}
The results for the decay rates are scheme-scale dependent.
The present analysis is restricted by
the one parametric family of the $MS$-type schemes \cite{MS}. The
ambiguity is due to the
dependence on the parameter $\mu$, which enters via dimensional regularization.
(For the similar analysis of the $\Gamma_{Z\rightarrow hadrons}$
see \cite{Sur}.) In the solution of the RG eq.(\ref{eq:RG})
one replaces $\mu_{\overline{MS}}^2=e^{t}M_H^2$ and
$\alpha_s \rightarrow \alpha_s^t$,  where the parametrization of
$\alpha_s^t$ is given by the eq.(\ref{eq:Asparametr})
with $\Lambda_t=e^{-t/2}\Lambda_{\overline{MS}}$.

\vspace{75mm}
\begin{center}
{\bf fig.5} The approximants of the
           $\Gamma_{H\rightarrow b\overline{b}}$ vs the scale parameter $t$.\\
\end{center}
\vspace{2mm}

   The one-, two- and three-loop approximants for the
$\Gamma_{H\rightarrow b\overline{b}}$ in terms of running quark mass
(eq.(\ref{eq:RGimpr}), with $N=5$ and $m_f=m_b$)
vs. the scale parameter $t$
 are plotted in the fig.5.
One can see that the higher order corrections diminish the scale dependence
from 40\% to nearly 5\%.
The solid curve (corresponding to the three-loop result) became flat
in the wide range of the logarithmic scale parameter $t$.
Moreover, the choice $t=0$
($\overline{MS}$-scheme) satisfies Stevenson's ``minimal sensitivity''
principle \cite{St1}. The mass corrections does not change
the overall picture. It should be stressed once again that the present
consideration is restricted by
the $MS$-type schemes. On the other hand one could  carry out the
scheme invariant analysis along the lines of \cite{St1,St2}
(see also \cite{PRV}).

   It is known that the dependence on the unphysical scale parameter
is due to the truncation of perturbation series at particular order.
If one sums up to ``all orders'', the scale dependence will disappear.
Thus, one may try to relate the remainder dependence on the
scale to the sum of all uncalculated terms and estimate
the size of the theoretical uncertainty in the reasonably wide neighborhood
of the initial choice of scale (in our case $t=0$). The deviation of
the approximant from the constant is used as a measure
of theoretical error. From the fig.5 one estimates the theoretical
uncertainty at 5\%.

    The comparison of fig.3 and fig.5 shows that the
high order corrections resolve the very large discrepancy
(more then factor 2!) between the results for
$\Gamma_{H\rightarrow b\overline{b}}$ in terms of running and
pole quark masses. The remaining discrepancy is about 5\%,
which is the same as the
estimated 5\% uncertainty due to the uncalculated higher order terms.

    Besides the above estimated theoretical error,  there are two more
contributions, with the potentially important effects comparable
with the present uncertainty. The first most likely non negligible
contribution comes from the diagrams in
fig.2, with the virtual top quark in the
triangle fermionic loop. (The similar diagrams in the case
of $Z$-boson decay gave sizable
correction \cite{Khn}.) The other contribution may come from the
diagrams in fig.1 with the virtual top quark. The decoupling theorem
may not apply in this case because of possible comparable size of
the top and the Higgs masses.
The contributions from these diagrams (and those in fig.2) with virtual
bottom and lighter quarks are already included in the result.
However, in the case of top, one needs to evaluate
these diagrams explicitly. The similar  diagrams for the $Z$-boson
decay gave a somewhat moderate but not entirely negligible correction
\cite{Sop}. The evaluation of the higher order mass corrections
($\sim m_f^6$ and higher) requires an accurate treatment
 of infrared divergences. On the other hand those corrections are heavily
suppressed by the powers of $m_f^2/M_H^2$ and numerically will
be negligible. The calculation of the $O(\alpha_s^3)$ corrections
requires the evaluation of relevant four-loop diagrams up to the
finite terms in their expansion
in Laurent series in $\varepsilon$ and the method used in \cite{Zde}
to simplify the $O(\alpha_s^3)$ calculations of the decay rate
of $Z\rightarrow hadrons$ does not apply. The above discussion of the
theoretical uncertainties
is mainly for the $b\overline{b}$ and $c\overline{c}$
modes. For the case of $H\rightarrow t\overline{t}$ the uncertainty
should be somewhat higher because of large mass corrections, unless
the Higgs mass is in the TeV region.

\vspace{4mm}
{\bf 7. Summary.}
 The quark mass corrections to the decay rate
of the SM Higgs particle into the quark-antiquark pair
 to $O(\alpha_s^2)$ perturbative QCD are calculated.
The previously known $O(\alpha_s)$ results with explicit mass
dependence
and the three-loop massless limit result with running quark mass
parametrization are independently confirmed.
The expression for the decay rate is obtained
both with running and pole quark mass parametrizations.
It was found that the quark mass
corrections are not entirely negligible and
they decrease the large and negative massless coefficients.
For the decay mode $H\rightarrow t\overline{t}$ the mass corrections are
large and important.
On the other hand, the higher order corrections resolve the large
discrepancy (more then factor 2) between the results for the decay width
in terms of running and pole quark masses.
It was also found that the
three-loop QCD correction reduces the scale dependence significantly.
The theoretical error of evaluation of the QCD
contribution was estimated
at 5\% for the decay mode $H\rightarrow b\overline{b}$ plus, the possible
effects of the virtual heavy top quark are emphasized.

  Finally one should mention the recent work \cite{Kat}, where the similar
problem without calculating three-loop mass corrections has been discussed.
However, the relation between running and pole quark masses used
in \cite{Kat}, seems to be incorrect.

\vspace{4mm}
{\bf Acknowledgements} The author is grateful to D.E.Soper for
discussions and the advice concerning the relation between
running and pole quark masses.
It is a pleasure to thank D.Broadhurst, B.Kniehl and A. Sirlin for helpful
communications and D.Strom for the discussion of the
experimental status of the problem at LEP.
This work was supported by the
U.S. Department of Energy under grant No. DE-FG06-85ER-40224.

\end{document}